# Non-uniform Thermal Conductivity in Nanoscale Multiple Hotspot Systems

Yu He,[1] Zhihao Zhou,[1,2] Lina Yang,[3,*] Nuo Yang[1,†]

[1]School of Science, National University of Defense Technology, Changsha 410073, China

[2]School of Energy and Power Engineering, Huazhong University of Science and Technology, Wuhan 430074, China

[3]School of Aerospace Engineering, Beijing Institute of Technology, Beijing 100081, China

*Contact author: yangln@bit.edu.cn

[†]Contact author: nuo@nudt.edu.cn

**Abstract**

Understanding nanoscale hotspot thermal transport is crucial in electronic devices. Contrary to common perception, recent experiments show that closely spaced nanoscale multiple hotspots can enhance heat dissipation. Here, the thermal transport in nanoscale multiple hotspot systems is investigated by solving the phonon Boltzmann transport equation. The local thermal conductivity is proposed to describe the non-uniform spatial distribution of heat transport capability in nanoscale multiple hotspot systems. The maximum value exceeds the uniform heating case by up to 27%, which is attributed to the spatially varying fraction of unscattered phonons emitted from hotspots. Moreover, the effects and mechanisms of hotspot spacing on thermal transport are investigated, showing that reducing the hotspot spacing can enhance the heat flux by up to 40%. This work challenges the conventional view that thermal transport capability is spatially uniform throughout the system and provides fundamental insights for thermal management in high-power-density integrated circuits.

## 1. Introduction

The rising power density of electronic devices, driven by advances in nanofabrication, intensifies the challenges of thermal management (1-3). These challenges are further compounded by heat transport mechanisms at the micro-nanoscale hotspots. When the hotspot size is reduced to or below the phonon mean free path (MFP), heat transport deviates from classical Fourier's law (4-9), leading to a suppression of thermal conductivity to values drastically below the bulk value (8-10). A fundamental understanding of these nanoscale phenomena is thus key to tackling thermal management in modern electronics.

Recent studies have revealed an anomalous heat transport phenomenon in systems with nanoscale multiple hotspots (11-13). Although nanoscale hotspots are conventionally viewed as impeding heat dissipation (14), extreme ultraviolet (EUV) scatterometry experiments show that nanoscale closely spaced multiple hotspots can instead enhance heat dissipation (15-18). Based on molecular dynamics (MD) simulations, Honarvar et al. (19) attribute this enhancement to phonon-phonon scattering between the nanoscale hotspots, which preferentially redirects phonons in the cross-plane direction. However, the system size in such MD simulations remains substantially smaller than experimental scales. The underlying physical mechanisms responsible for the enhanced heat dissipation over broader scales remain poorly understood, particularly given the discrepancies among different experimental measurements (18, 20).

These open questions underscore a more fundamental need for accurate characterization of thermal transport capability in nanoscale multiple hotspot systems. At nanoscale, thermal conductivity as defined by Fourier's law becomes invalid due to its strong dependence on system size (21-23) and hotspot size (7, 14). While the effective thermal conductivity is widely used as a black box measure of overall thermal transport capability (1, 4), it inherently assumes that thermal transport capability remains uniform across different spatial locations. However, recent studies suggest that thermal transport capability can be spatially non-uniform (24, 25), particularly near interfaces (26), boundaries (27, 28) and nanoscale hotspots (29-32). A quantitative description of this spatial variation remains challenging, and existing approaches cannot fully resolve the spatial distribution of thermal transport capability.

In this study, the thermal transport in systems with nanoscale multiple hotspots was investigated by solving the phonon Boltzmann transport equation (BTE). First, the local thermal conductivity was proposed to characterize the spatial distribution of thermal transport capability. Second, the local thermal conductivity was calculated and compared across configurations with a nanoscale single hotspot, nanoscale multiple hotspots, and uniform heating conditions. Finally, systems with varying hotspot spacing and substrate thickness were studied, and the mechanism for the enhanced thermal transport capability with reduced hotspot spacing was analyzed.

## 2. Methods

Over the past few decades, the phonon BTE has been demonstrated to be a reliable method for describing heat transport at the micro-nanoscale (33-36). In this work, the phonon BTE is solved using the Monte Carlo method (37-40). Silicon is selected as the substrate material due to its prevalence in chips. The interatomic force constants for silicon are calculated using VASP (41) and serve as inputs for ShengBTE (42) to obtain phonon properties. The phonon properties, including group velocity, relaxation time, and heat capacity, then serve as inputs for the Monte Carlo simulations (43).

The fin field-effect transistor (FinFET) array, commonly found in integrated circuits, contains nanoscale periodic hotspots due to the system's structure and electrical characteristics. A schematic illustration of a FinFET array and different heating configurations are presented in Fig. 1. In the simulations, the hotspot regions are set as isothermal boundaries at 310 K. The regions between hotspots are set as adiabatic boundaries. The heat sink regions are set as isothermal boundaries at 300 K. All other boundaries are treated as periodic boundaries. For comparison, in all periodic hotspot cases investigated in this study, the hotspot region size S and the spacing L between hotspots are kept equal. Additional simulation details and validation are provided in the Supporting Material.

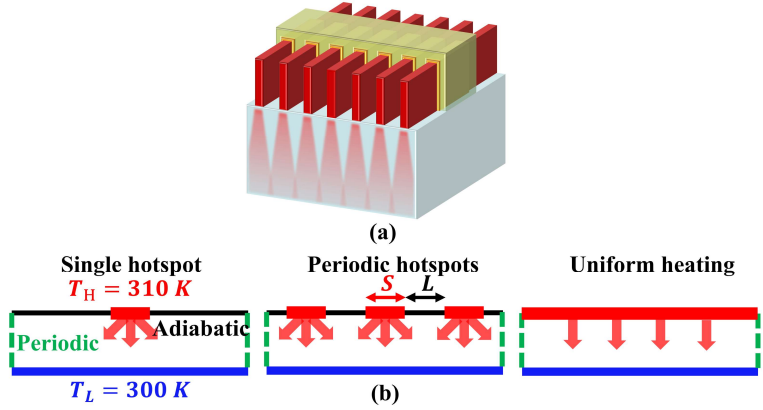

**Fig. 1.** (a) Schematic illustration of a FinFET array. (b) Schematic diagrams of the single-hotspot, periodic-hotspot, and uniform heating systems. The red regions represent the hotspots of size S, maintained at 310 K as isothermal boundaries. The black regions represent the adiabatic boundaries between hotspots with spacing L. The blue region represents the heat sink, set to 300 K as isothermal boundaries. The green dashed line represents the periodic boundaries.

Several definitions of thermal conductivity have been proposed at the micro-nanoscale since the macroscopic Fourier's definition is invalid (25). A common approach defines thermal conductivity as $k = q/[(T_H - T_L)/D]$. When $T_H$ and $T_L$ correspond to the temperatures of the hotspot and the heat sink, respectively, and D is the distance between them, the resulting value is termed the effective thermal conductivity (4, 44). Alternatively, when $T_H$ and $T_L$ are measured within the intermediate linear temperature region near the hotspot and heat sink, respectively, and D represents the length of that linear region, the calculated thermal conductivity is referred to as the intermediate linear region thermal conductivity (11, 19). Here, $q$ denotes the heat flux in the direction from the hotspot to the heat sink. For simplicity, the thermal conductivity in the intermediate linear temperature region under uniform heating conditions is specifically denoted as $k_u$ in this study.

In nanoscale multiple hotspot systems, thermal transport becomes multidirectional and there is no a simple linear temperature distribution (30). Since the thermal transport capability varies with spatial location and cannot be accurately represented by a single thermal conductivity value, a local thermal conductivity is proposed to characterize the spatial dependence, defined as:

$$\boldsymbol{k}_{loc}(x,y) = \frac{\boldsymbol{q}(x,y)}{|\nabla T(x,y)|} \quad (1)$$

That is, the local thermal conductivity is a vector whose direction aligns with that of the heat flux vector $\boldsymbol{q}$. In classical Fourier's law, the heat flux and temperature gradient are collinear but opposite. At the micro-nanoscale, however, ballistic phonons can carry heat flux without contributing to the temperature gradient, causing the two vectors to deviate from collinearity. The direction of the heat flux becomes more critical, as it incorporates the contribution of ballistic phonons.

The concept of local thermal conductivity provides a spatially resolved description of thermal transport capability, which challenges classical Fourier's law. In the diffusive regime, thermal conductivity is treated as an intrinsic material property, independent of spatial location. This framework cannot capture non-diffusive transport, whereas the local thermal conductivity, as calculated using the phonon BTE, naturally incorporates the contribution of ballistic and quasi-ballistic phonons.

Furthermore, this concept differs fundamentally from the conventional effective thermal conductivity (4, 14). While the effective thermal conductivity offers a global, system-averaged scalar that obscures internal spatial variations, the spatial distribution of $k_{loc}$ reveals microscopic details of heat transport, thereby providing a more accurate description of localized heat dissipation enhancement or localized thermal resistance increase. This establishes local thermal conductivity as an essential concept for understanding and designing non-uniform thermal systems at the nanoscale.

## 3. Results and discussion

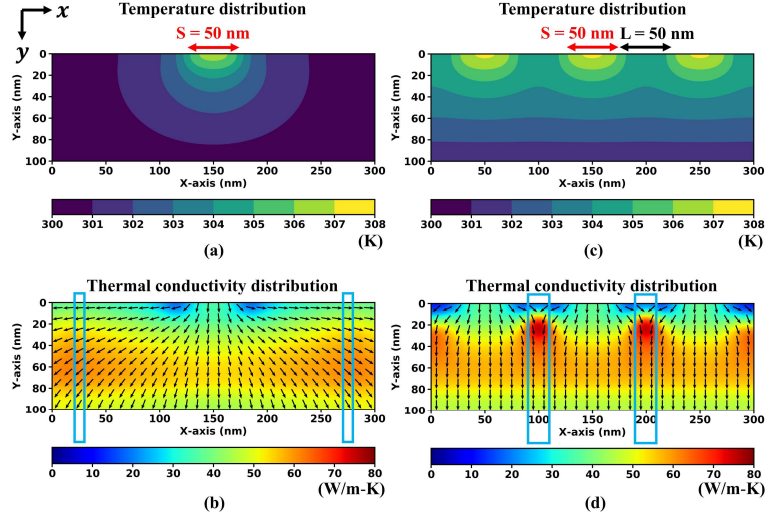

**Fig. 2.** Distribution of temperature and thermal conductivity for a substrate thickness of D = 100 nm. (a) Temperature distribution and (b) thermal conductivity distribution for a nanoscale single hotspot. (c) Temperature distribution and (d) thermal conductivity distribution for nanoscale periodic hotspots. Black arrows in (b) and (d) indicate the heat flux direction. The blue box marks the region of maximum local thermal conductivity, which will be analyzed in Fig. 3. For the nanoscale single hotspot system, a total x-direction length of 1200 nm was used to ensure adequate separation from the boundaries, with the central 300 nm region selected for comparison and a hotspot size S = 50 nm. In the system with nanoscale periodic hotspots, both the hotspot size S and the spacing L are 50 nm.

The temperature distribution for a nanoscale single hotspot is presented in Fig. 2(a). A distinct temperature jump is observed at the boundary positions, that is, a significant temperature reduction near the hotspot location, which is consistent with previous studies (4, 45). It should be noted that in this single hotspot simulation, the total system length in the x-direction was set to 1200 nm, a sufficiently large distance to ensure that the hotspot region is unaffected by the periodic boundary conditions.

The thermal conductivity distribution for a nanoscale single hotspot is presented in Fig. 2(b). In this case, the region of maximum local thermal conductivity is located away from the hotspot, with a value reaching 62.2 W/(m·K), which equals $k_u$, the thermal conductivity under uniform heating conditions. Except at the periodic boundaries, the local thermal conductivity near other boundaries is reduced due to phonon-boundary scattering.

The temperature distribution for nanoscale periodic hotspots is presented in Fig. 2(c). Compared with the single hotspot case, the temperature jump is reduced by 0.7 K under periodic hotspot conditions, indicating that the boundary temperatures are closer to the hotspot temperature.

The thermal conductivity distribution for nanoscale periodic hotspots is presented in Fig. 2(d). In this case, the maximum local thermal conductivity is found in the interstitial regions between hotspots, enhanced to a value of 79.3 W/(m·K). This value is greater than those in both single hotspot and uniform heating conditions.

A key finding of this study is the spatial distribution of thermal transport capability. As observed in Fig. 2(b) and (d), the local thermal conductivity cannot be described by a single value for the entire system, but rather manifests as a spatial distribution that characterizes the spatially dependent thermal transport capability. This spatial variation has been corroborated by several experiments (26, 28, 29), even though it is often overlooked when using the oversimplified single-value representation.

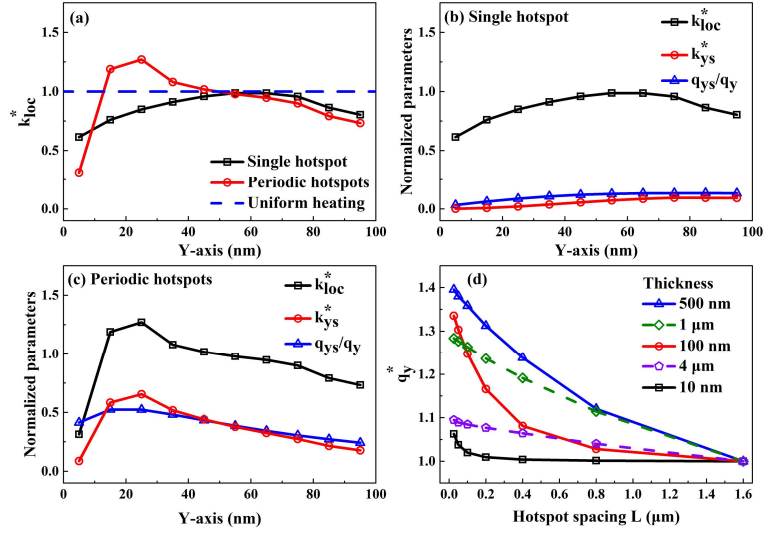

**Fig. 3.** (a) Comparison of normalized local thermal conductivity $k_{loc}^*$ in the blue box region for single and periodic hotspot systems. (b) Normalized local thermal conductivity $k_{loc}^*$, normalized local thermal conductivity of unscattered phonons $k_{ys}^*$ and contribution of unscattered phonons to heat flux $q_{ys}/q_y$ in the maximum local thermal conductivity region, with the corresponding region indicated by the blue box in Fig. 2(b) for the single hotspot system. (c) The $k_{loc}^*$, $k_{ys}^*$ and $q_{ys}/q_y$ in the maximum local thermal conductivity region, with the corresponding region indicated by the blue box in Fig. 2(d) for the periodic hotspot system. (d) Dimensionless heat flux in the y-direction $q_y^*$ as a function of hotspot spacing L for substrate thicknesses D of 10 nm, 100 nm, 500 nm, 1 μm, and 4 μm. The local thermal conductivity is normalized as $k_{loc}^* = k_{loc}/k_u$. The normalized local thermal conductivity of unscattered phonons is calculated by $k_{ys}^* = (q_{ys}/|\nabla T|)/k_u$. The heat flux is normalized as $q_y^* = q_y/q_{min}$, performed independently for each thickness, where $q_{min}$ is the minimum value for each respective thickness. In all simulated periodic hotspot cases, the hotspot region size S and the spacing L between hotspots are maintained equal, collectively occupying 50% of the top region.

For direct comparison, Fig. 3(a) displays the regions of maximum value of $k_{loc}$ for the nanoscale single hotspot, nanoscale periodic hotspot, and uniform heating configurations. It can be observed that the maximum value of $k_{loc}$ in the single hotspot case cannot exceed $k_u$, whereas in the periodic hotspot case it exceeds $k_u$ by 27%. This enhancement of local thermal conductivity reveals the presence of non-diffusive thermal transport in periodic hotspot systems.

To further clarify the enhancement of the maximum value of $k_{loc}$, the contribution of phonons emitted from the hotspot before undergoing scattering events (referred to as unscattered phonons) to the heat flux is analyzed. The rate of their contribution is quantified by $q_{ys}/q_y$, where $q_{ys}$ is the heat flux contributed by unscattered phonons and $q_y$ is the total heat flux in the y-direction. For direct comparison with $k_{loc}$, the local thermal conductivity corresponding to unscattered phonons is calculated by $k_{ys} = q_{ys}/|\nabla T|$.

As shown in Fig. 3(b) and (c), nanoscale single- and periodic-hotspot systems exhibit distinct behaviors. Compared with the single hotspot system, the periodic hotspot system shows strong similarity between the trends of its $k_{ys}^*$ and $k_{loc}^*$ curves. Additionally, $q_{ys}/q_y$ reaches only about 0.1 for the single hotspot system while exceeding 0.5 for the periodic hotspot system. These observations indicate that the enhancement of local thermal conductivity in periodic hotspot systems is attributed to the increased fraction of unscattered phonons. This occurs because when the hotspot spacing is smaller than the phonon MFP, unscattered phonons emitted from adjacent hotspots can propagate into the central region between hotspots, thereby significantly increasing their fraction in these regions. Conversely, as the hotspot spacing increases (eventually approaching the single-hotspot case when sufficiently large), the fraction of unscattered phonons in the central region between hotspots decreases, since these phonons undergo scattering during propagation. In regions where the fraction of unscattered phonons is higher, the transport behavior is closer to that dominated by ballistic transport. Therefore, this spatial variation of the fraction of unscattered phonons leads to the spatial non-uniformity in thermal transport capability.

To investigate the effect of hotspot spacing on thermal transport in nanoscale periodic hotspot systems, Fig. 3(d) presents the heat flux as a function of hotspot spacing for different substrate thicknesses. As the hotspot spacing L decreases (with both L and S decreasing simultaneously while being kept equal), the heat flux along the y-direction increases—however, this enhancement shows a clear thickness dependence. At a thickness D = 10 nm, the enhancement is only 10%, but it grows with increasing thickness, reaching 40% at a thickness D = 500 nm. Further increasing the thickness beyond this value reduces the enhancement. This thickness-dependent enhancement arises from the variation in the fraction of phonons absorbed by hotspots at different thicknesses, as will be discussed in detail in Fig. 4. In summary, the heat flux is enhanced with reduced hotspot spacing, and the extent of this enhancement first increases and then decreases with increasing substrate thickness.

To analyze the mechanism of enhanced heat flux with reduced hotspot spacing, phonon transport processes are tracked. As shown in the schematic representation in Fig. 4(a), the case where hotspot spacing L (equal to hotspot size S) is much larger than the phonon MFP is considered. After phonons are emitted from a hotspot, they propagate toward the heat sink. During this propagation, phonon-phonon scattering may occur, randomly altering phonon directions. Some phonons continue moving toward the heat sink (positive y-direction), while others are redirected back toward the hotspot (negative y-direction). These phonons returning toward the hotspot are critical for thermal transport. Given the limited phonon MFP, phonons are nearly unable to reach the boundaries adjacent to the hotspots, as the distance is too far comparing with the phonon MFP. Consequently, almost all phonons returning toward the hotspots are absorbed by the hotspots.

The typical case with hotspot spacing L = 1.6 μm is analyzed in Fig. 4(c), where L is much larger than the phonon MFP. The heat flux from phonons absorbed by the hotspot is denoted $q_a$, while that from phonons scattered by nearby boundaries without being absorbed is denoted as $q_{un}$. The results show that most phonons moving toward the hotspot (negative y-direction) are absorbed, with only a low fraction remaining unabsorbed.

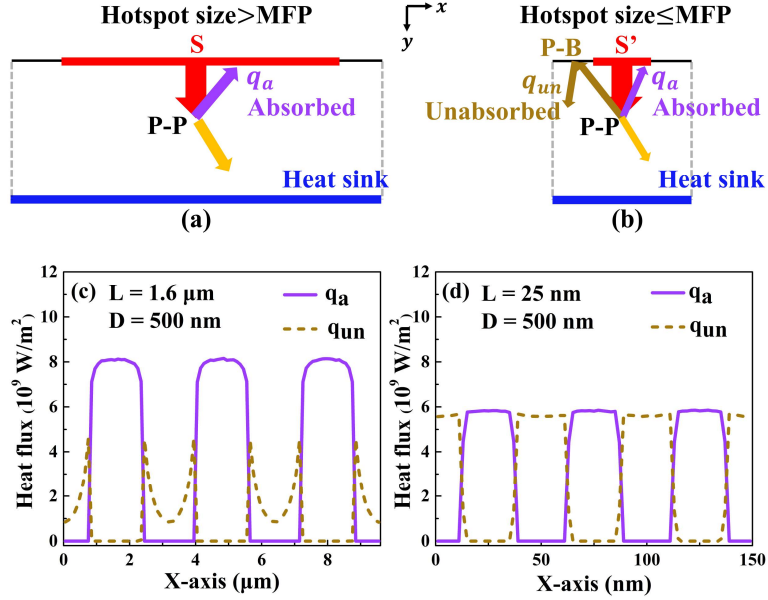

**Fig. 4.** (a) Schematic of phonon transport in nanoscale periodic hotspot systems with spacing L (equal to hotspot size S) much larger than the phonon MFP. (b) Schematic of phonon transport with spacing L (equal to hotspot size S) comparable to or smaller than the phonon MFP. The hotspot and spacing regions each occupy 50% of the top region. Phonon-phonon and phonon-boundary scattering are denoted by P-P and P-B, respectively. The absorbed and unabsorbed heat fluxes by hotspots are denoted by $q_a$ and $q_{un}$, respectively. (c) Distribution of heat flux absorbed by hotspots and heat flux not absorbed by hotspots along the x-axis for hotspot spacing L = 1.6 μm and (d) L = 25 nm at a substrate thickness D = 500 nm.

As shown in the schematic representation in Fig. 4(b), the case where hotspot spacing L (equal to hotspot size S) is comparable to or smaller than the phonon MFP is considered. Here, phonons can reach boundaries near hotspots. This leads to an increased probability of phonon-boundary scattering and a reduced probability of phonon absorption by hotspots. Those phonons scattered by the boundary may continue to propagate toward the heat sink, thereby enhancing the heat flux along the y-direction.

The typical case with hotspot spacing L = 25 nm is analyzed in Fig. 4(d), where L is smaller than the phonon MFP. It can be observed that among phonons moving toward the hotspot (negative y-direction), approximately half are absorbed by the hotspot while the other half undergo phonon-boundary scattering. This reduction in absorbed heat flux increases the heat flux reaching the heat sink, thereby enhancing the thermal transport capability.

As previously discussed, heat flux enhancement from reduced hotspot spacing results from a decreased fraction of phonons being absorbed by hotspots. However, this fraction is also influenced by the substrate thickness. When the substrate thickness is much smaller than the phonon MFP, phonon transport is nearly ballistic, and most phonons do not return to the hotspots. When the substrate thickness is much larger than the phonon MFP, sufficient space is provided for phonon-phonon scattering, which enables phonons to propagate to distant boundaries. In both scenarios, the fractions of absorbed and unabsorbed phonons are already nearly equal, leaving limited room for a further reduction in the absorption fraction and thereby limiting the potential heat flux enhancement. As a result, the heat flux enhancement weakens when the substrate thickness is either significantly smaller or larger than the phonon MFP.

The direct experimental mapping of local thermal conductivity remains challenging, primarily due to the difficulty in resolving the heat flux field at the nanoscale. Although recent advances in scanning transmission electron microscopy electron energy-loss spectroscopy (STEM-EELS) have enabled sub-nanometer temperature resolution (26, 46), measuring local heat flux in periodic hotspots is still formidable. Alternatively, indirect validation can be pursued by comparing the measured temperature gradient fields, obtained via techniques such as STEM-EELS or tip-enhanced Raman thermal measurement (6, 29), with those predicted by phonon BTE simulations.

## 4. Conclusions

Here, the thermal transport in systems with nanoscale multiple hotspots was systematically investigated by solving the phonon BTE. It was demonstrated that the local thermal conductivity in nanoscale multiple hotspot systems exhibit significant spatial variations. The maximum value in multiple hotspots exceeds the value in uniform heating due to an increased fraction of unscattered phonons emitted from hotspots. Furthermore, a reduction in hotspot spacing is shown to enhance heat flux by up to 40%, resulting from reduced phonon absorption by hotspots. This study reveals the long-overlooked spatial distribution of thermal transport capability in nanoscale, and provides broad implications for thermal management in high-power-density integrated circuits.

# Supplementary Information: Non-uniform Thermal Conductivity in Nanoscale Multiple Hotspot Systems

Yu He,[1] Zhihao Zhou,[1,2] Lina Yang,[3,*] Nuo Yang[1,†]

[1]School of Science, National University of Defense Technology, Changsha 410073, China

[2]School of Energy and Power Engineering, Huazhong University of Science and Technology, Wuhan 430074, China

[3]School of Aerospace Engineering, Beijing Institute of Technology, Beijing 100081, China

*Contact author: yangln@bit.edu.cn

†Contact author: nuo@nudt.edu.cn

# 1. Monte Carlo method

The computational methodologies commonly employed for investigating micro- and nanoscale heat conduction encompass the phonon Boltzmann transport equation (BTE) and molecular dynamics simulations. While molecular dynamics simulations provide detailed atomistic information, their prohibitive computational cost renders them infeasible for systems extending to hundreds of nanometers. Consequently, the solution of the phonon BTE, is regarded as an advanced approach that balances accuracy with computational tractability. The Monte Carlo (MC) method is utilized to solve the phonon BTE, expressed as:

$$\frac{\partial f}{\partial t} + \boldsymbol{V}_g \cdot \nabla f = \frac{f^{loc} - f}{\tau} \tag{S1}$$

where $f$ denotes the phonon distribution function, and $f^{loc}$ represents the equilibrium Bose-Einstein distribution at the local temperature. Additionally, $\boldsymbol{V}_g$ and $\tau$ correspond to the phonon group velocity and relaxation time, respectively.

To enhance computational efficiency and reduce statistical variance, particularly under small deviations from equilibrium, the deviational formulation is introduced. The governing equation is recast in terms of deviational energy (1):

$$\frac{\partial e^d}{\partial t} + \boldsymbol{V}_g \cdot \nabla e^d = \frac{\left(e^{loc} - e^{eq}_{T_{eq}}\right) - e^d}{\tau} \tag{S2}$$

The deviational energy distribution is defined as $e^d = e - e^{eq}_{T_{eq}}$. Here, $e = \hbar\omega f$, $e^{loc} = \hbar\omega f^{loc}$ and $e^{eq}_{T_{eq}} = \hbar\omega f^{eq}_{T_{eq}}$, with $f^{eq}_{T_{eq}} = \left[exp(\hbar\omega/k_b T_{eq}) - 1\right]^{-1}$ being the Bose-Einstein distribution at the equilibrium temperature $T_{eq}$, and $k_b$ is Boltzmann's constant.

Further simplification is achieved through linearization, which is valid when $|T - T_{eq}| \ll T_{eq}$. The linearized form of the deviational BTE is given by (2):

$$\frac{\partial e^d}{\partial t} + \boldsymbol{V}_g \cdot \nabla e^d = \frac{\mathcal{L}(e^d) - e^d}{\tau} \tag{S3}$$

where

$$\mathcal{L}(e^d) = \left(T_{loc} - T_{eq}\right) \frac{de^{eq}_{T_{eq}}}{dT} \tag{S4}$$

The solution of the linearized deviational energy-based BTE (LBTE) using a Monte Carlo approach involves generating phonon bundles sampled from the initial deviational energy distribution. These phonon bundles are propagated through drift and scattering processes governed by the LBTE.

For steady-state simulations, thermodynamic observables such as temperature deviation and heat flux are computed by time-averaging phonon bundle trajectories. In a region of volume $V$, the temperature deviation from equilibrium $T_{dev}$ and the heat flux component $q_y$ are evaluated using (3):

$$T_{dev} = \frac{\dot{\mathcal{E}}^d_{eff}}{CV} \sum_i s_i \frac{l_i}{V_{g,i}} \tag{S4}$$

$$q_y = \frac{\dot{\mathcal{E}}^d_{eff}}{V} \sum_i s_i l_{y,i} \tag{S5}$$

where $s_i$ signifies the sign of the deviational energy (positive or negative), $l_i$ is the total path length traveled by the i-th phonon bundle, $l_{y,i}$ is its displacement along the y-direction, $C$ is the volumetric specific heat. Here, $\dot{\varepsilon}_{eff}^d = \dot{E}_{tot}^d/N$ is the effective deviational energy rate, and $\dot{E}_{tot}^d$ is the total deviational energy rate, $N$ is the number of phonon bundles in the simulation. Further details regarding the Monte Carlo method can be found in previous literature (1-4).

The heat flux in the y-direction can be decomposed based on the category of phonon bundles as follows:

$$q_y = q_{ys} + q_{ypp} + q_{ypb} \tag{S6}$$

In practical calculations, the decomposed heat fluxes are obtained by separately applying the summation in Equation S5 to different categories of phonon bundles. Phonons emitted from the heat source that have not undergone any scattering are referred to as unscattered phonons, and the heat flux generated by these unscattered phonons is denoted as $q_{ys}$. Phonons that have been scattered are classified: those having undergone phonon-phonon scattering are referred to as phonon-phonon (P-P) scattered phonons, with their associated heat flux being $q_{ypp}$, while those having undergone phonon-boundary scattering are referred to as phonon-boundary (P-B) scattered phonons, with their associated heat flux being $q_{ypb}$. The purpose of decomposing the heat flux is to quantify the strength of various phonon effects. A high $q_{ys}$ indicates a large fraction of unscattered phonons and a strong ballistic transport. Details on the validation of the simulation accuracy are provided in previous studies (3, 5).

## 2. Computational models

For the single nanoscale hotspot case, the computational model is shown in Fig. S1(a). The system parameters are set with hotspot size S = 50 nm, substrate thickness D = 100 nm, and system length LA = 1200 nm. This length ensures that heat transport near the hotspot is unaffected by x-direction boundaries. In regions far from the hotspot where the temperature exactly equals the ambient temperature and the temperature gradient is 0, the local thermal conductivity cannot be calculated. Therefore, a 300 nm region centered on the hotspot is selected for calculating distributions. For calculating the temperature distribution and thermal conductivity distribution, the numbers of meshes in the x-, y-, and z-directions are set to 30, 10, and 1, respectively. In the other regions, no mesh is generated, which improves the computational speed. In these unmeshed regions, phonon bundles can move, but the temperature and heat flux are not calculated. All boundaries in the z-direction are set to periodic for all simulation cases.

For the periodic nanoscale hotspot case, the computational model is shown in Fig. S1(b). Two different computational setups are employed in the various periodic nanoscale hotspot simulations. When the focus is on the internal thermal transport mechanisms, the thermal conductivity distribution needs to be calculated. In such cases, the parameters used are hotspot size S = 50 nm, spacing L = 50 nm, substrate thickness D = 100 nm, system length LB = 300 nm, with the numbers of meshes in the x-, y-, and z-directions set to 30, 10, and 1, respectively.

When focusing on the overall heat dissipation of the system rather than local regions, calculating the heat flux is more direct than obtaining the thermal conductivity distribution. This heat flux indicates how much heat is transferred from the hotspots to the heat sink. To investigate the overall heat dissipation at different hotspot spacings, the heat flux values under various spacing conditions are calculated. In periodic nanoscale hotspot systems, the total heat flow in the y-direction is conserved. Consequently, the mean heat flux remains the same when calculated at different cross-sections along the y-axis. To quantify the phonon bundles moving in the negative y-direction that are absorbed or unabsorbed by hotspots, the mean heat flux in the top 1 nm thick region of the computational domain is calculated. This value is denoted as $q_y$.

In all periodic nanoscale hotspot simulations, the hotspot size S and the spacing L between hotspots are maintained equal, collectively occupying 50% of the top region respectively. This configuration ensures that both S and L are varied simultaneously when adjusting the hotspot spacing.

For the uniform heating case, the computational model is shown in Fig. S1(c). The system is configured with length LC = 100 nm and substrate thickness D = 100 nm. Given the identical results at different x-positions under uniform heating, the numbers of meshes in the x-, y-, and z-directions are set to 1, 10, and 1, respectively. The thermal conductivity in the central linear temperature region $k_u$ is calculated for comparison, as commonly employed in previous studies (6, 7).

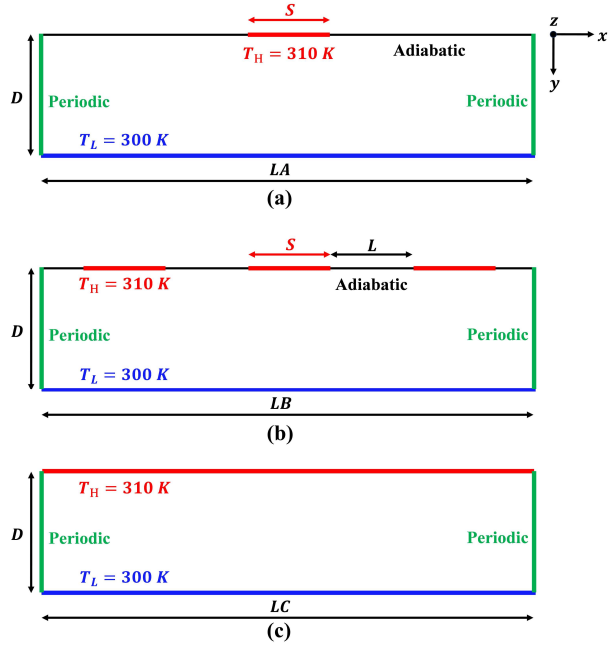

**Fig. S1.** Computational models for (a) the single nanoscale hotspot system, (b) the multiple nanoscale hotspot system, and (c) the uniform heating system.

## 3. Computational convergence

Prior to simulation, the convergence of key results with respect to the number of phonon bundles and the maximum number of scattering events was verified. Insufficient phonon bundles lead to significant statistical errors, while an inadequate maximum number of scattering events prevents the system from reaching steady state.

As shown in Fig. S2(a), to ensure the system reaches steady state, the convergence of $q_y$ with the maximum number of scattering events was tested for the largest system. The largest system is characterized by a length LB = 9.6 μm and a substrate thickness D = 4 μm. Convergence is observed when the maximum number of scattering events reaches $1 \times 10^5$. Accordingly, this value is adopted for all simulations.

Fig. S2(b) shows the convergence of $q_y$ with the number of phonon bundles for the largest system, ensuring sufficiently small statistical errors. Convergence is achieved with $1 \times 10^7$ phonon bundles.

For thermal conductivity distribution calculations, where smaller statistical errors are required,

the convergence of the maximum local thermal conductivity $k_{loc}$ with the number of phonon bundles was separately tested, as shown in Fig. S2(c). Based on these convergence tests, $1 \times 10^8$ phonon bundles are used in thermal conductivity distribution calculations.

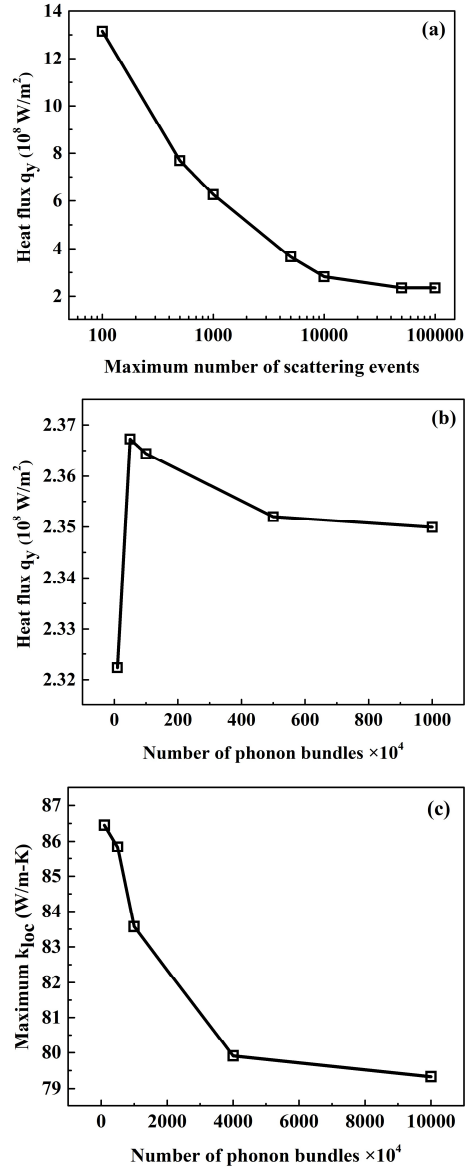

**Fig. S2.** (a) Convergence of the y-direction heat flux $q_y$ with the maximum number of scattering events. (b) Convergence of the y-direction heat flux $q_y$ with the number of phonon bundles. (c) Convergence of the maximum local thermal conductivity $k_{loc}$ with the number of phonon bundles.